\documentclass{SCIS}

\newcommand \ssr {Space Sci. Rev.}
\newcommand \aap { Astron. Astrophys.}
\newcommand \apj {Astrophys. J.}
\newcommand \apjl {Astrophys. J. Lett.}
\newcommand \apjs {Astrophys. J. Suppl.}
\newcommand \solphys {Sol. Phys.}
\newcommand \lrsp {Living Rev. Sol. Phys.}
\newcommand \araa {Ann. Rev. Astron. Astrophys.}
\newcommand \jgr {Journal of Geo. Res. (Space Physics)}
\newcommand \raa {Research in Astronomy and Astrophysics}

\begin{document}
\ensubject{subject}
\ArticleType{Article}
\Year{2022}
\Month{January}
\Vol{65}
\No{3}
\DOI{10.1007/s11433-021-1836-y}
\ArtNo{239611}
\ReceiveDate{November 18, 2021}
\AcceptDate{December 15, 2021}
\OnlineDate{January 19, 2022}

\title{Persistent fast kink magnetohydrodynamic waves detected in a quiescent prominence}{Persistent fast kink magnetohydrodynamic waves detected in a quiescent prominence}

\author[1,2]{Dong Li}{{lidong@pmo.ac.cn}}
\author[1]{Jianchao Xue}{}
\author[3]{Ding Yuan}{{yuanding@hit.edu.cn}}
\author[1,4]{Zongjun Ning}{}

\AuthorMark{D. Li}

\AuthorCitation{D. Li, J. Xue, D. Yuan, and Z. Ning}

\address[1]{Key Laboratory of Dark Matter and Space Astronomy, Purple Mountain Observatory, Chinese Academy of Sciences, Nanjing {\rm 210023}, China}
\address[2]{CAS Key Laboratory of Solar Activity, National Astronomical Observatories, Beijing {\rm 100101}, China}
\address[3]{Institute of Space Science and Applied Technology, Harbin Institute of Technology, Shenzhen, Guangdong {\rm 518055}, China}
\address[4]{School of Astronomy and Space Science, University of Science and Technology of China, Hefei {\rm 230026}, China}

\abstract{Small-scale, cyclic, transverse motions of plasma threads
are usually seen in solar prominences, which are often interpreted
as magnetohydrodynamic (MHD) waves. Here, we observed small-scale
decayless transverse oscillations in a quiescent prominence, and
they appear to be omnipresent. The oscillatory periods of the
emission intensity and a proxy for the line-of-sight Doppler shift
are about half period of the displacement oscillations. This feature
agrees well with the fast kink-mode waves in a flux tube. All the
moving threads oscillate transversally spatially in phase and exhibit
no significant damping throughout the visible segments, indicating
that the fast kink MHD waves are persistently powered and ongoing
dissipating energy is transferred to the ambient plasma in the quiet
corona. However, our calculations suggest that the energy taken by
the fast kink MHD waves alone can not support the coronal heating on
the quiet Sun.}



\maketitle

\begin{multicols}{2}
\section{Introduction}\label{sec1}
Solar prominences are cool and dense elongated plasma structures
embedded in the surrounding hot corona above the solar limb. Those
dark prominences detected on the solar disk are called solar
filaments. Usually, the prominence plasmas are about one hundred
times cooler and denser than their surrounding coronal plasmas,
which raise an important issue for their origin and stability
\cite{Mackay10,Gibson18}. As one of the most surprising structures
suspended by coronal magnetic fields, solar prominences are often
\Authorfootnote regarded as the source/driver of solar eruptions,
especially for the large-scale eruption, such as the solar flare or
coronal mass ejection \cite{Parenti14,Chen20}. High-resolution
observations suggest that they typically consist of a large number
of dynamic thread-like structures
\cite{Lin05,Okamoto07,Okamoto15,Ning09a,Yan15,Bi20}, and such
thread-like structures also exist in the galaxy clusters
\cite{Zhou20}. These dynamic behaviors are often linked to the
magnetic activities in the corona
\cite{Li13,Schmieder13,Ning09b,Shen15,Taroyan19}, and are therefore
employed to understand the origin and physical properties of solar
prominences. According to the relative locations, solar prominences
are usually divided into active region and quiescent prominences.
The active region prominences refer to cooler plasmas suspended
above the strong magnetic field region, and they are often highly
dynamic and relatively short-lived \cite{Okamoto07,Okamoto15}.
On the contrary, the quiescent prominences are often seen in the
quiet-Sun regions, more frequently at high solar latitudes. They are
more stable and usually have a longer lifetime
\cite{Ning09a,Shen15,Bi20}.

Prominences often show oscillatory motions with a wide range of
periods from several minutes to tens of minutes and even several
hours, which have been confirmed by the ground- and space-based
observations \cite{Engvold01,Foullon04,Pouget06,Zhang17,Arregui18}.
According to the oscillation amplitude, prominence oscillations can
be classified as large-scale or small-scale \cite{Oliver02}. The
large-scale prominence oscillation could disturb a large volume of
the whole prominence. This motion is often triggered by the external
disturbances, such as Moreton waves or Extreme ultraviolet (EUV)
waves, solar flares, and jets \cite{Asai12,Shen14,Zhang20}. The
small-scale prominence oscillation is usually detected as the
oscillatory motion of a fine thread. It only affects a small part of
the prominence. The prominence oscillation could be regarded as a
self-oscillation of the coronal plasma structure
\cite{Martin08,Ning09a,Nakariakov16,Li18}, or it may represent a
magnetohydrodynamic (MHD) wave mode
\cite{Okamoto07,Okamoto15,Lin09,Morton12} that reveals the local
magnetic nature. The interactions between the fluid motion
and magnetic field convert the dynamic energy into smaller length
scales, and fine magnetic dynamics could even dissipate the energy
into kinetic scale, which are the very nature of MHD waves. On the
other hand, the MHD waves are much more general than the
wave-guiding effect, and they may carry enough energies which could
be converted into heat. Thus, the MHD waves are regarded as a
potential source for the plasma heating on the Sun or the Sun-like
stars \cite{Gibson18,Van20,Melis21,Srivastava21}.

A number of papers reported small-scale transverse
oscillations in the plasma threads of solar prominences
\cite{Okamoto07,Lin07,Lin09,Li18}. Those transverse MHD waves are
thought to be driven in the photosphere and propagate upwards along
the magnetic field line into the prominence embedded in the corona
\cite{Nakariakov20,Melis21}. The transverse oscillation with the
same phase was observed in the moving thread of an active region
prominence \cite{Okamoto07}. However, they failed in determining the
exact oscillatory mode, largely due to the absence of line-of-sight
(LOS) Doppler velocity measurements. In this paper, we identify that
the fast kink mode waves should be responsible for those small-scale
transverse oscillations in a quiescent prominence, based on the
H$\alpha$ and its LOS velocity measurements from the New Vacuum
Solar Telescope (NVST) \cite{Liu14,Yan20}.

\section{Observations}\label{sec2}
On 08 December 2016, the NVST captured a sequence of high-resolution
images of a quiescent prominence above the north-west solar limb,
i.e., N21W89. The recording lasted for about 4.5 hours, ranging from
04:00~UT to 08:31~UT. The observed data can be accessed at the NVST
website \textsf{http://fso.ynao.ac.cn/cn/datashow.aspx}. The
H$\alpha$ images at wavelengths of 6562.8~{\AA} and its two
off-bands at $\pm$0.3~{\AA} are used in this study. And the level-1
images are used, which are reconstructed by the frame selection from
a large number of raw images by the NVST team \cite{Liu14,Yan20}.
They have a spatial scale of about 0.165~arcsec per pixel, and a
time cadence of about 40~s at each bandpass. The data of the
Atmospheric Imaging Assembly aboard the Solar Dynamics Observatory
(SDO/AIA) at the wavelength of 193~{\AA} are also used here, which
have been calibrated with the standard routines in the Solar
SoftWare package \cite{Lemen12}. It has a time cadence of 12~s, and
each pixel corresponds to 0.6~arcsec. It is mainly used to refer the
accurate location of the quiescent prominence.

In this study, the ground-based NVST images at different wavebands
are co-aligned by the local cross-correlation technique using the
FLCA code\footnote{$https://github.com/xuejcak/flca$} \cite{Xue21},
which is an efficient method based on the Fourier local correlation
tracking \cite{Fisher08}. The co-alignment between three NVST
wavebands were done by referencing to the AIA 193 {\AA} images.
Here, the NVST H$\alpha$ images at the different bands can be
aligned as precise as one pixel. Therefore, the LOS velocity
measurement is derived directly from two nearly simultaneous NVST
H$\alpha$ off-band images at two extended wings \cite{Li18,Yan20},
i.e., $\pm$0.3~{\AA}. Noting that the LOS velocity is a qualitative
description rather than a quantitative calculation, it is not the
speed value of Doppler shifts.

\begin{figure}[H]
\centering
\includegraphics[width=0.45\textwidth]{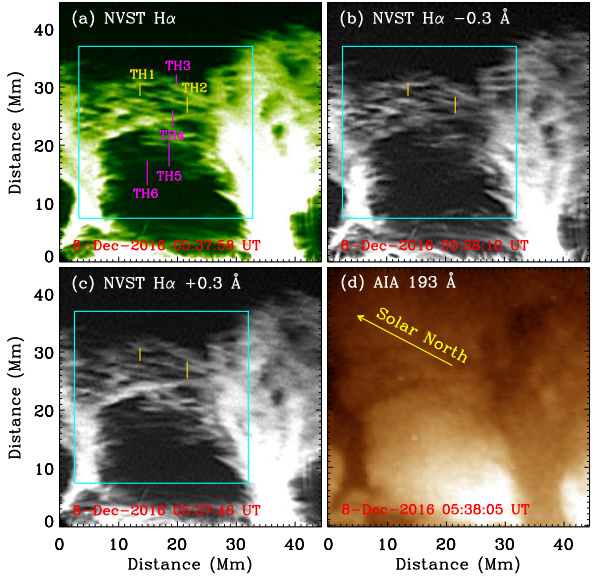}
\caption{Overview of the quiescent prominence on 08 December 2016.
Field-of-view measured in NVST H$\alpha$ at the line center (a), and
two extended wings at $\pm$0.3~{\AA} (b, c), and AIA 193~{\AA} (d).
The short vertical lines indicate the positions of moving threads of
interest, and gold lines are shown in the main text, while the
magenta lines are displayed in the Appendix figures. The cyan box
marks the region used to perform the animation. The yellow arrow in
panel~(d) indicated the solar north. \label{image}}
\end{figure}

\cref{image} shows a bright quiescent prominence suspended above the
solar limb in H$\alpha$ line center (a), and two extended wings at
$\pm$0.3~{\AA} (b, c), respectively. The quiescent prominence
consists of two bright barbs that are connected by a great number of
thread-like structures as seen in the NVST H$\alpha$ images. The
prominence threads are very thin, similar to the moving threads seen
in the Ca II H line measured by the Hinode/SOT \cite{Okamoto07}. On
the other hand, the double bright barbs are perpendicular to the
solar limb, and they are darker than the surrounding corona in
AIA~193~{\AA} due to the continuum photoionization, as shown in
\cref{image}~(d). It should be pointed out that the solar prominence
has been rotated in \cref{image}, so that the moving threads are
horizontal. We cannot find the fine-scale threads in AIA~193~{\AA},
largely due to its low spatial resolution. The NVST animation
(S1.mp4) reveals continuous horizontal motions nearly parallel to
the solar limb between two bright barbs of the prominence. These
moving threads appear to undergo oscillatory motions in the
plane-of-the-sky, which is mainly perpendicular to the moving
threads. To investigate the small-scale transverse oscillatory
motions in the quiescent prominence, we choose six isolated moving
threads that are almost parallel to the solar surface, the thread
positions are marked by the short lines in \cref{image}~(a).

\section{Data reductions and Results}
\label{sect3}
In order to measure the oscillatory periods and
amplitudes of these transverse motions in the moving threads, we
first select a crosscut slit that is placed perpendicular to the
axis of one moving thread, and then the time-distance plot is
constructed along the fixed cut slit. Next, the brightest pixels
along the transverse oscillation are selected manually. Finally, a
sine function with a linear background (i.e., Eq.~\ref{eq1}) is
applied to fit the transverse oscillation.
\begin{equation}
  A(t)=A_m \sin(\frac{2 \pi}{P}~t+ \psi )+kt+A_0.
\label{eq1}
\end{equation}
where $A_m$ represents the displacement amplitude of the transverse
oscillation and $P$ is the oscillatory period. While $\psi$ is the
initial phase of the oscillatory motion, $A_0$ represents the
position around which the plasma thread oscillates, and $k$ is the
drifting velocity in the plane-of-the-sky \cite{Li18,Zhang20}. With
the derivative of Eq.~\ref{eq1} we can get the velocity amplitude,
i.e., $v=2\pi A_m/P$, as ref \cite{Lin09}.

\begin{figure}[H]
\centering
\includegraphics[width=0.45\textwidth]{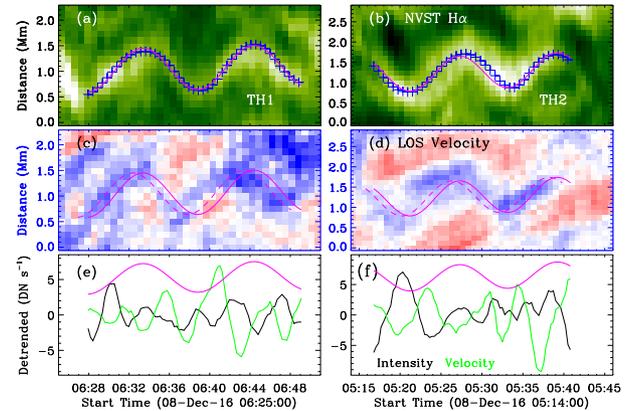}
\caption{Panels~(a) \& (b): Time-distance images taken in the NVST
H$\alpha$ bandpass along the perpendicular slits marked by two gold
lines in \cref{image}. The blue pluses (`+') mark the bright pixels
of the oscillating threads, whereas the continuous magenta line
draws a sinusoidal fit. Panels~(c) \& (d): Time-distance images
taken in the LOS Doppler velocity along the same two cut slits. The
dashed magenta line shows the thread motion on LOS velocity
observation, for comparison with the fitted motion (continuous
magenta line) in panels~(a) \& (b). Panels~(e) \& (f): Detrended
time series of the emission intensity (black) and Doppler velocity
(green) measured afloat on the oscillatory positions (`+').}
\label{tds}
\end{figure}

Figures~\ref{tds}~(a)$-$(d) draw the time-distance images along the
moving threads~1 and 2 in the H$\alpha$ bandpass and its
line-of-sight (LOS) velocity, both threads show small-scale
displacement oscillations for at least one cycle. Their oscillation
amplitudes are estimated to be around 420~km and 410~km without any
significant damping, and the periods are about 11.0~minutes and
12.0~minutes, respectively. Table~\ref{tab} also presents the
velocity amplitude ($v$) of each sample measurement.
Figures~\ref{tds}~(c) \& (d) show that the oscillatory motions tend
to be blue-shifted and they reach the maximum (i.e., dashed curves)
a bit earlier. Here, the time delay between the H$\alpha$ line
center and LOS velocity are obtained by shifting the fit function
with time through the cross-correlation analysis, as shown by the
dashed and solid magenta lines. The phase shift is the time lag
between the transverse displacement of the oscillating thread in the
plane-of-the-sky and the LOS Doppler shift oscillatory signal, but
it is normalized to the oscillation period. The short time delay
suggests a small phase shift of the transverse oscillation between
the H$\alpha$ line center and its LOS velocity.

To quantify the displacement oscillations of the prominence fine
threads, we extract the raw time series from the oscillatory
positions (blue pluses) for the H$\alpha$ emission intensity and the
LOS velocity, and then obtain the trend time series with a smooth
window of about 20~minutes. At last, the detrended time series are
calculated by subtracting the trended time series from the raw time
series. Figures~\ref{tds}~(e) \& (f) present the quantitative
analysis of the H$\alpha$ emission intensity (black) and Doppler
shift (green) measured afloat on the oscillatory positions. We can
see at least four peaks in the Lagrangian emission intensity and
Doppler shift, which are twice as more than the double peaks of the
displacement oscillation during the same time interval. This
suggests that the oscillatory periods in the Lagrangian emission
intensity and Doppler shift are no more than half period of the
displacement oscillations. On the other hand, the NVST H$\alpha$
animation (S1.mp4) suggests that there is not a violent or obvious
eruption nearby the moving threads in the quiescent prominence.
However, we shall note that it might occur because that the NVST
instrument is not sensitive enough to capture the fine-scale
eruptive features, which are as small as the NVST observational
limitation.

\begin{figure}[H]
\centering
\includegraphics[width=0.45\textwidth]{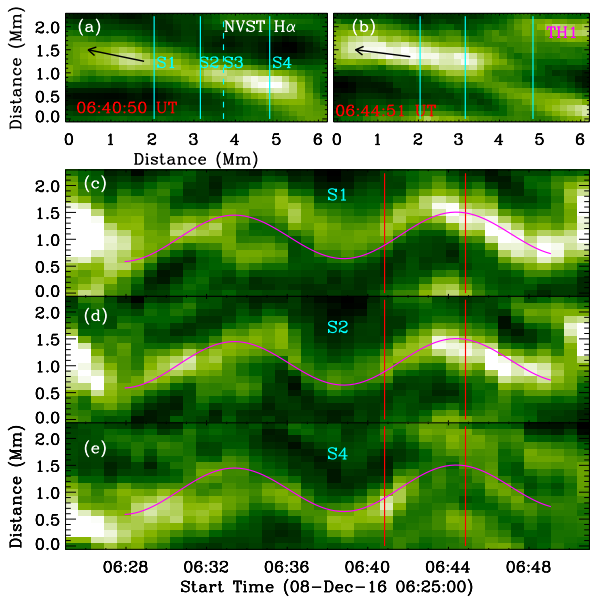}
\caption{Oscillations at multiple positions of the thread TH1.
Panels~(a) \& (b): Two snapshots in H$\alpha$ line center at two
fixed instances of time. The vertical cyan lines outline the slit
positions, noting that the dashed line corresponds to TH1 in
\cref{image}, and the arrows indicate the moving direction of the
prominence thread. Panels~(c)$-$(e): Time-distance plots
corresponding to different perpendicular slits, as marked by the
solid cyan lines in panel~(a). Two vertical red lines mark the
thread time shown in panels~(a) \& (b). \label{slit1}}
\end{figure}

\begin{figure}[H]
\centering
\includegraphics[width=0.45\textwidth]{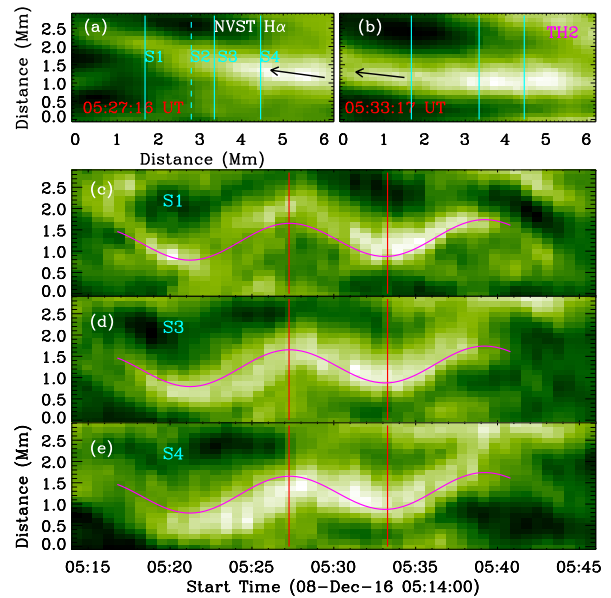}
\caption{Similar to \cref{slit1} but for the prominence thread TH2.
\label{slit2}}
\end{figure}

To reach the conclusion that each moving thread oscillates
transversally with the same phase, a given thread with several cut
slits along its length is considered, as shown in
Figures~\ref{slit1} and \ref{slit2}. \cref{slit1} presents the
snapshots and time-distance images of the plasma thread~1 (TH1)
along another three different cut slits, for instance, before and
after the fixed cut slit. Panels~(a) \& (b) show two snapshots of
TH1 in NVST H$\alpha$ line center at 06:40:50~UT and 06:44:51~UT,
respectively. Four cut slits are made along the moving thread and
their time-distance images are plotted in panels~(c)$-$(e), as
indicated by S1$-$S4. Note that the cut slit 3 (dashed line) is at
the same position as TH1 in \cref{image}~(a), and the corresponding
time-distance image is given in \cref{tds}~(a), where the magenta
curve is obtained. The similar displacement oscillation with the
same period can be found in these cut slits along the entire length
of the moving thread in NVST H$\alpha$ bandpass, and no apparent
time delay is seen among them, suggesting that there is almost no
phase shift along the moving thread for the transverse displacement
oscillation. Moreover, they all appear to be decayless. The
constant-amplitude oscillations mean that the dissipative losses are
compensated by the energy supply. \cref{slit2} shows the snapshots
(a \& b) of TH2 at two fixed instances of time and the time-distance
image (c$-$e) corresponding to three different perpendicular slits.
Similar to the transverse oscillation in the moving thread TH1, the
displacement oscillation at three positions along the moving thread
TH2 also have an identical period, and there is no significant time
delay between them, as shown by the magenta curve in \cref{slit2}.
All those observational facts suggest that the small-scale
transverse oscillations are most likely to be persistent in the
quiescent prominence. The supplementary files such as
Figures~\ref{slit3}$-$\ref{slit6} present the other four isolated
threads (TH3$-$TH6) and their transverse displacement oscillations
at different cut slits, further confirming our results. That is, the
transverse displacement oscillations can be found at any sites along
the moving thread with the same period, without any significant time
delay and damping, implying the omnipresence of small-scale
decayless transverse MHD waves.

\section{Conclusion and Discussion} \label{sec4}
The small-scale transverse oscillations are largely perpendicular to
the moving threads found in the quiescent prominence. Key parameters
measured in the six moving threads are summarized in
Table~\ref{tab}. The displacement amplitudes range from 420~km to
980~km, whereas the respective velocity amplitudes are
4.0~km~s$^{-1}$ and 6.2~km~s$^{-1}$. The oscillation periods are
measured to be about 10.3$-$16.5~minutes, these values are larger
than those observed in the moving threads of an active region
prominence (about 2$-$4~minutes) \cite{Okamoto07}, but similar to
the case in quiescent prominences \cite{Zhou18}. On the other hand,
we can not detect any eruptive events near the moving threads during
the oscillatory time, suggesting that those small-scale transverse
oscillations could be regarded as the self-oscillatory processes of
the magnetic structure in the quiescent prominence
\cite{Martin08,Lin09,Ning09a,Nakariakov16,Li18} rather than being
induced by external disturbances \cite{Asai12,Shen14,Zhang20}.
Recent study \cite{Afanasyev20} found that the random motion could
excite decayless transverse oscillations in coronal loops, which
might be used to explain the excitation of transverse oscillatory
motions seen in the quiescent prominence, since such a behavior
appears without apparent impulsive energy release or periodic
driving.

\begin{table*}[ht]
\footnotesize
\caption{Key parameters measured in six moving threads.}
\label{tab}      
\tabcolsep 15pt  
\begin{tabular*}{\textwidth}{cccccccc}
\bottomrule
   Notation                 & Parameter                            & TH1    & TH2    &  TH3   &  TH4   &  TH5   &  TH6  \\\hline
displacement amplitude      &  $A_m$ (km)                          & 420    & 450    &  510   &  420   &  980   &  720  \\
oscillatory period          &  $P$ (minutes)                       & 11.0   & 12.0   & 12.1   & 10.3   & 16.5   & 12.6  \\
velocity amplitude          &  $v$ (km~s$^{-1}$)                   & 4.0    & 3.9    &  4.4   & 4.3    & 6.2    &  6.0  \\
minimum wavelength          & $\lambda_{min}$ ($\times$10$^{3}$km) & 109    & 121    & 127    & 106    & 153    & 121   \\
phase speed                 &  $c_{ph}$ (km~s$^{-1}$)              & 165    & 168    & 175    & 172    & 155    & 160   \\
Alfv\'{e}n speed            &  $V_{A}$ (km~s$^{-1}$)               & 117    & 119    & 124    & 122    & 110    & 113   \\
magnetic field strength     &  $B$ ($\times$10$^{-4}$T)            & 5.4    & 5.5    & 5.7    & 5.6    & 5.0    & 5.2   \\
magnetic field perturbation &  $b$ ($\times$10$^{-4}$T)            & 0.131  & 0.129  & 0.144  & 0.139  & 0.201  & 0.195 \\
energy flux density         &   $<E_k>$ (W~m$^{-2}$)               & 16.7   & 16.2   & 21.4   & 19.9   & 37.3   & 36.2  \\
\bottomrule
\end{tabular*}
\end{table*}

An intriguing feature is that the transverse displacement
oscillations measured at any positions are mostly perpendicular to
the moving threads, there is no apparent time delay in different cut
slits along the entire length of each moving thread. That is, the
transverse oscillation appears to oscillate synchronously along the
entire length of each moving thread. It seems that the transverse
oscillations always exist (or persist) in the quiescent prominence,
when an isolated thread moves to an appropriate position and
direction, i.e., the moving thread is largely perpendicular to the
oscillation and no other threads or bright patches are mixed with
it, then the transverse oscillation can be seen in the plane of the
sky. Magnetic fields in the solar/stellar corona are often thought
to play a major role in guiding MHD waves. Therefore, the transverse
oscillations observed in moving threads of the quiescent prominence
are most likely to be regarded as transverse MHD waves in the corona
\cite{Nakariakov20}, such as fast kink waves
\cite{Van08,Terradas08,Lin09}, or Alfv\'{e}n waves \cite{Melis21}.
The transverse MHD waves are persistent in the quiescent prominence,
which are similar to the persistent kink/Alfv\'{e}n waves observed
in coronal loops \cite{Tian12,Anfinogentov13}. The omnipresence of
decayless transverse oscillations of coronal loops have also been
demonstrated to be a common phenomenon in the solar corona
\cite{Anfinogentov15}, while both the coronal loops and prominence
threads could be regarded as the thin magnetic flux tubes in the
corona \cite{Diaz01,Okamoto07,Goossens13}. On the other hand, to
maintain the coronal temperature at one million K in the quiet Sun,
there should be interplay between continuous cooling and heating
processes \cite{Kolotkov21}. Since the small-scale transverse MHD
waves are omnipresent in the quiet corona, they could provide
ongoing energy input that balances the energy losses in the
solar/stellar quiet corona \cite{Tian14,Van20,Srivastava21}.

If we further consider the LOS velocity, the moving threads undergo
similar transverse oscillatory motions with a very close period at
the blue-shifted wing, and the phase shifts of
$\sim$$\frac{1}{5}\pi-\frac{1}{4}\pi$ are detected. Moreover, half
period of transverse displacement oscillations can be found in the
time series taken from the oscillatory positions at the H$\alpha$
emission intensity and Doppler shift (\cref{tds}~e \& f), which
agrees well with the forward modeling of MHD waves in the fast kink
mode of magnetic flux tubes such as prominence fine threads
\cite{Joarder97,Diaz01} or coronal loops \cite{Yuan16a,Yuan16b}.
Therefore, the small-scale transverse oscillations observed in
moving threads appear to be more appropriately explained as the fast
kink mode waves that propagate along the magnetic field lines in the
corona on the quiet Sun, and they might play a certain role in
heating the quiet corona \cite{Okamoto07,Melis21}. However, it is
still unknown if the fast kink MHD waves convey enough energy to
heat the quiet corona here.

The kink-like motions in prominence threads have been well
documented \cite{Arregui18}. The Doppler velocities along a filament
thread were found to oscillate with the same phase and period
\cite{Yi91}, and the Doppler velocities were measured over a
rectangular area. Moreover, they \cite{Yi91} reported the
longitudinal oscillation along the plasma threads rather than
oscillations normal to the magnetic field flux. Using the high
spatial resolution observations, the kink-like mode waves were
detected as the transverse displacement oscillations along the
perpendicular slit in filament/prominence threads, which were
excited by EUV waves or solar flares \cite{Liu12,Xue14}. The kink
oscillations were also observed simultaneously along the LOS and in
the plane-of-the-sky \cite{Lin09,Okamoto15}. Moreover, The
small-scale transverse oscillations were seen in an active region
prominence, and each moving thread oscillates transversally in phase
with the same period \cite{Okamoto07}. However, authors
\cite{Okamoto07} were not able to determine the exact oscillatory
mode, due to the lack of Doppler velocity observations. Later
numerical calculations \cite{Van08} suggested that the kink-mode
waves should be responsible for those thread oscillations. In this
study, we observed the omnipresence of small-scale transverse
oscillations in a quiescent prominence. Each plasma thread shows
transverse oscillations with the same phase and period, while the
time series of emission intensity and Doppler velocity reveal only
half period of the displacement oscillations. All our observational
results indicate that the small-scale transverse oscillation could
be regarded as the fast kink-mode MHD wave
\cite{Joarder97,Diaz01,Yuan16a,Yuan16b}.

Based on the kink wave model in magnetic flux tube
\cite{Okamoto07,Lin09,Yuan16a,Yuan16b,Nakariakov21}, the magnetic
field strengths in the quiescent prominence are then estimated to be
about (5.0$-$5.7)$\times$10$^{-4}$~T, which are in accordance with
previous findings in fine thread-like structures of solar
filament/prominence on the quiet Sun
\cite{Lin09,Schmieder13,Shen14}. The Lagrangian perturbations of
magnetic fields are also estimated using the Lagrangian displacement
vector \cite{Yuan16a}, which are roughly equal to
(0.129$-$0.201)$\times$10$^{-4}$~T. The Alfv\'{e}n speeds are
estimated to be 110$-$124~km~s$^{-1}$, similar to the measurement
(100~km~s$^{-1}$) in quiescent prominences \cite{Arregui18}.
However, the magnetic field strengths and Alfv\'{e}n speeds are much
smaller than those measured in the moving threads of the active
region prominence \cite{Okamoto07}, which are estimated to be
roughly 5$\times$10$^{-3}$~T and $>$1050~km~s$^{-1}$, respectively.
We shall note that our measurements are the lower limit estimations
using the prominence seismology.

To make clear if the energy taken by the fast kink MHD waves is
enough to heat the quiet corona, the time-averaged energy flux
density ($<E_k>$) \cite{Morton12} for the fast kink MHD waves is
then estimated in the quiescent prominence, which is situated in the
corona on the quiet Sun, as shown in Table~\ref{tab}. The energy
flux densities in the six moving threads are estimated to be about
16.2$-$37.3~W~m$^{-2}$, which are not sufficient for heating the
quiet corona, i.e., 100$-$200~W~m$^{-2}$, as reported in Ref.
\cite{Withbroe77,Aschwanden07}. It is commonly accepted that the
energy flux density calculated by the expression for fast kink MHD
waves (i.e, Eq.~\ref{eq5}) could be overestimated \cite{Goossens13}.
Therefore, the real energy flux taken by the persistent kink MHD
wave is not efficient to heat the quiet corona, although they could
provide ongoing heating. Our result is consistent with previous
findings, for instance, the observed MHD waves do not have enough
energy to heat solar active regions \cite{Klimchuk06,Klimchuk15}.

We want to stress that the periods of small-scale transverse motions
detected in the quiescent prominence are about 10.3$-$16.5~minutes,
which are similar to the oscillatory periods of about 5$-$16 minutes
observed in two quiescent filaments \cite{Yi91}. The similar periods
of 3$-$9~minutes are also reported in the individual thread of a
quiescent filament, which were regarded as the evidence of traveling
waves \cite{Lin07}. All those observed periods can be grouped into
the short-period (1$-$20~minutes) category \cite{Engvold01}, and
they could be regarded as the decayless MHD waves propagating along
the fine threads \cite{Lin07}. On the other hand, the animation
(S1.mp4) shows that the bright barbs undergo large-scale
perturbations. However, the period of large-scale perturbations
seems to be much larger than those periods of the transverse motions
in fine threads. Because the animation (S1.mp4) is from about 05:10
UT to 07:50 UT, including all the moving threads in our study, the
examined transverse motions are most likely to be the kink-mode MHD
waves adhering to individual threads \cite{Lin07,Okamoto07} rather
than part of the collective behavior of the entire prominence
system.

\section{Summary}\label{sec5}
In this study, we report the small-scale transverse oscillatory
motions in six moving threads of a quiescent prominence measured by
the NVST in H$\alpha$ and its LOS velocity images. They are
identified as the persistent fast kink-mode waves, without
significant damping. Our observations indicate that a balance could
be maintained between the wave energy dissipation and injection in
the quiet corona. However, the wave energy carried by the fast kink
MHD waves alone is not sufficient for heating the coronal plasma at
the quiet Sun. It is believed the forthcoming 1.8-meter telescopes
\cite{Rao20} can detect more detailed features of this type of
small-scale oscillations in filaments/prominences.

\Acknowledgements {We acknowledged two anonymous referees for their
valuable suggestions and inspiring comments. The authors thank the
NVST and SDO/AIA teams for providing the data. This study is
supported by NSFC under grant 11973092, 12173012, 12111530078,
12073081, U1631242, 11820101002, 11790302, U1731241, and the CAS
Strategic Priority Research Program on Space Science, Grant No.
XDA15052200, XDA15320103, and XDA15320301. D.L. is also supported by
the CAS Key Laboratory of Solar Activity (KLSA202003) and the
Surface Project of Jiangsu Province (BK20211402). D.Y. is supported
by the Shenzhen Technology Project (GXWD20201230155427003-20200804151658001). The Laboratory No. is
2010DP173032.}

\InterestConflict{The authors declare that they have no conflict of interest.}


\begin{appendix}
\setcounter{figure}{0} \makeatletter
\renewcommand{\thefigure}{A\@arabic\c@figure}

\section{Prominence oscillations in other four moving threads}
\begin{figure}[H]
\centering
\includegraphics[width=0.45\textwidth]{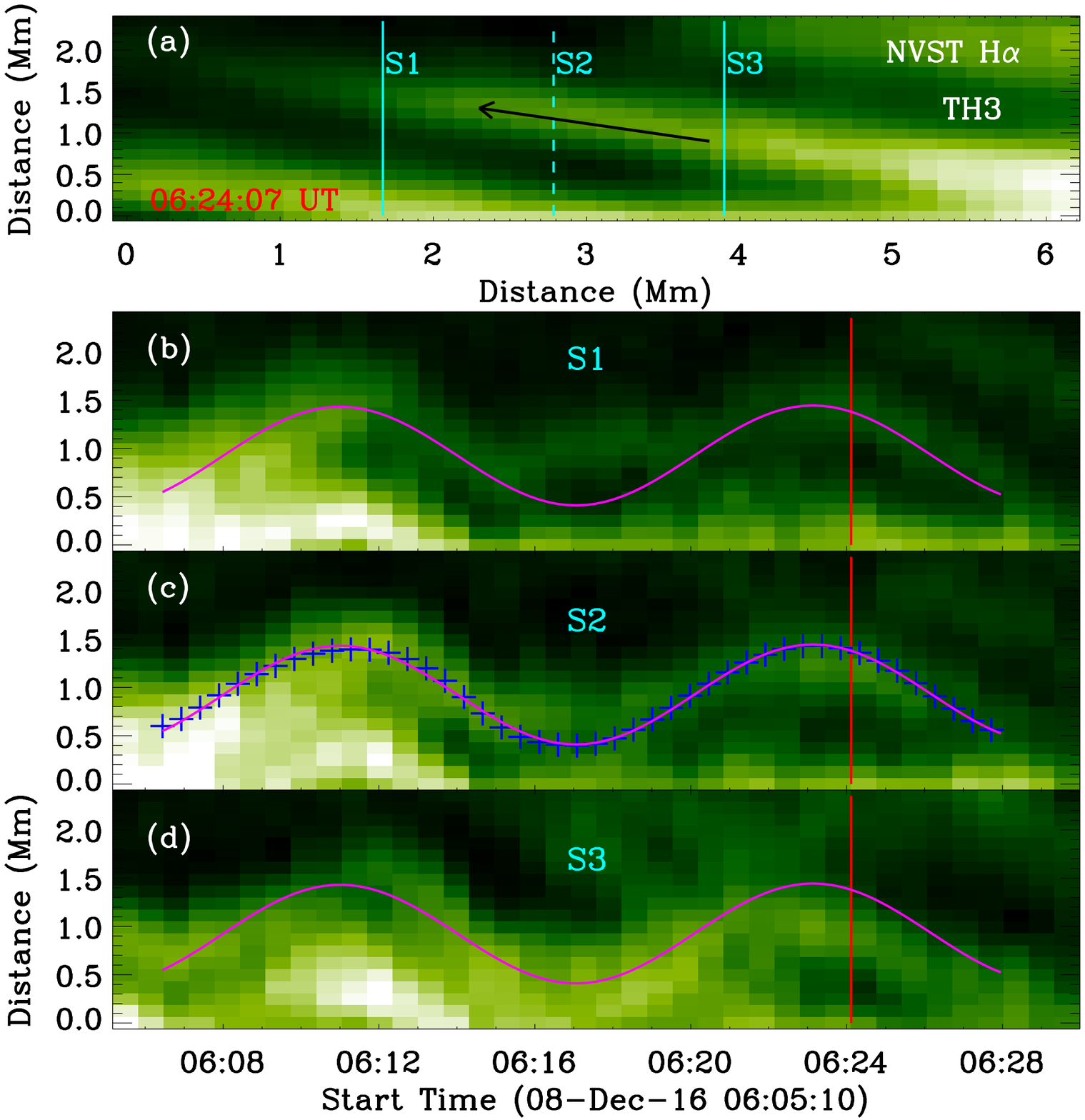}
\caption{Similar to \cref{slit1} but for the prominence thread TH3.
\label{slit3}}
\end{figure}

\begin{figure}[H]
\centering
\includegraphics[width=0.45\textwidth]{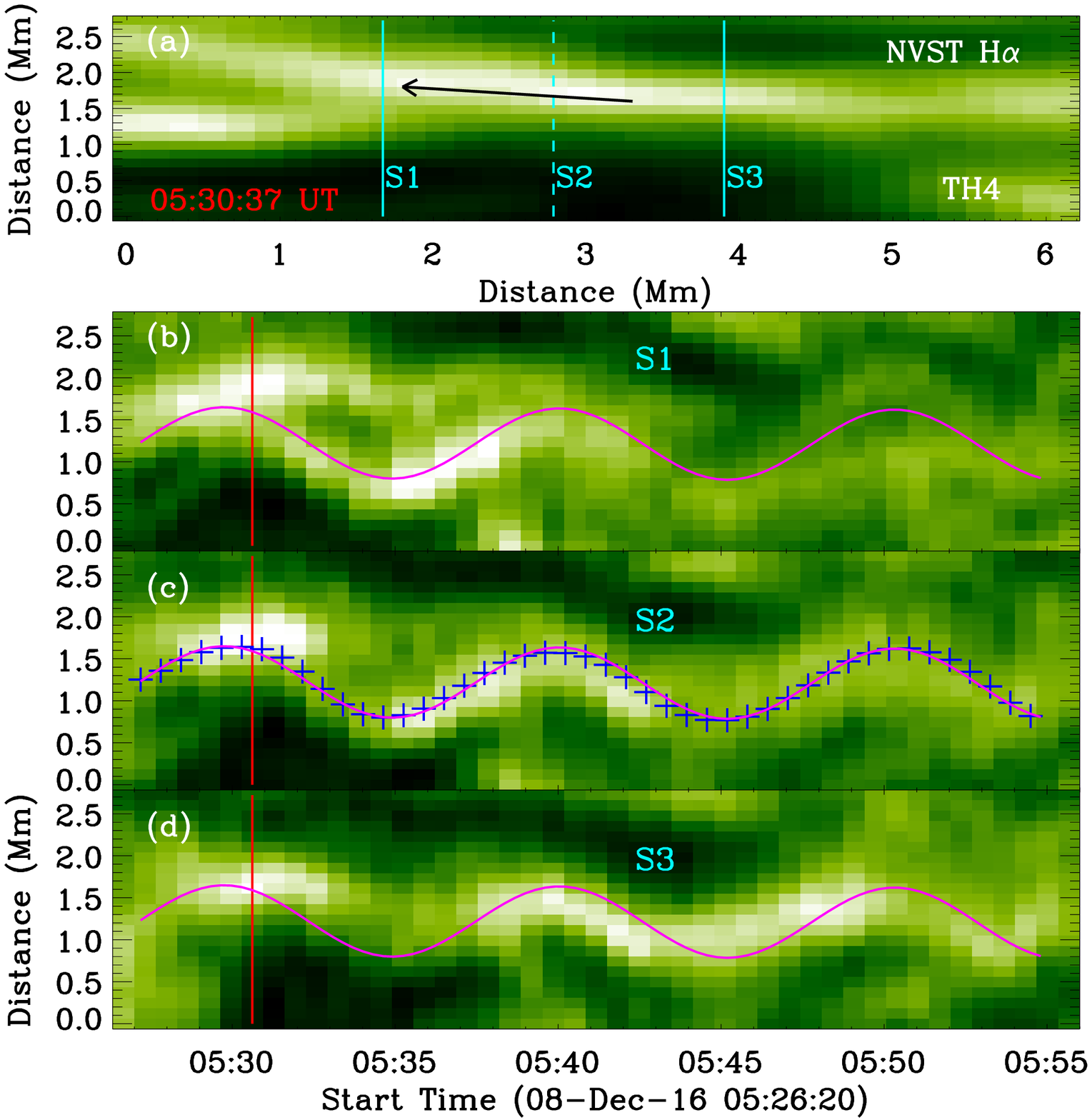}
\caption{Similar to \cref{slit1} but for the prominence thread TH4.
\label{slit4}}
\end{figure}

\begin{figure}[H]
\centering
\includegraphics[width=0.45\textwidth]{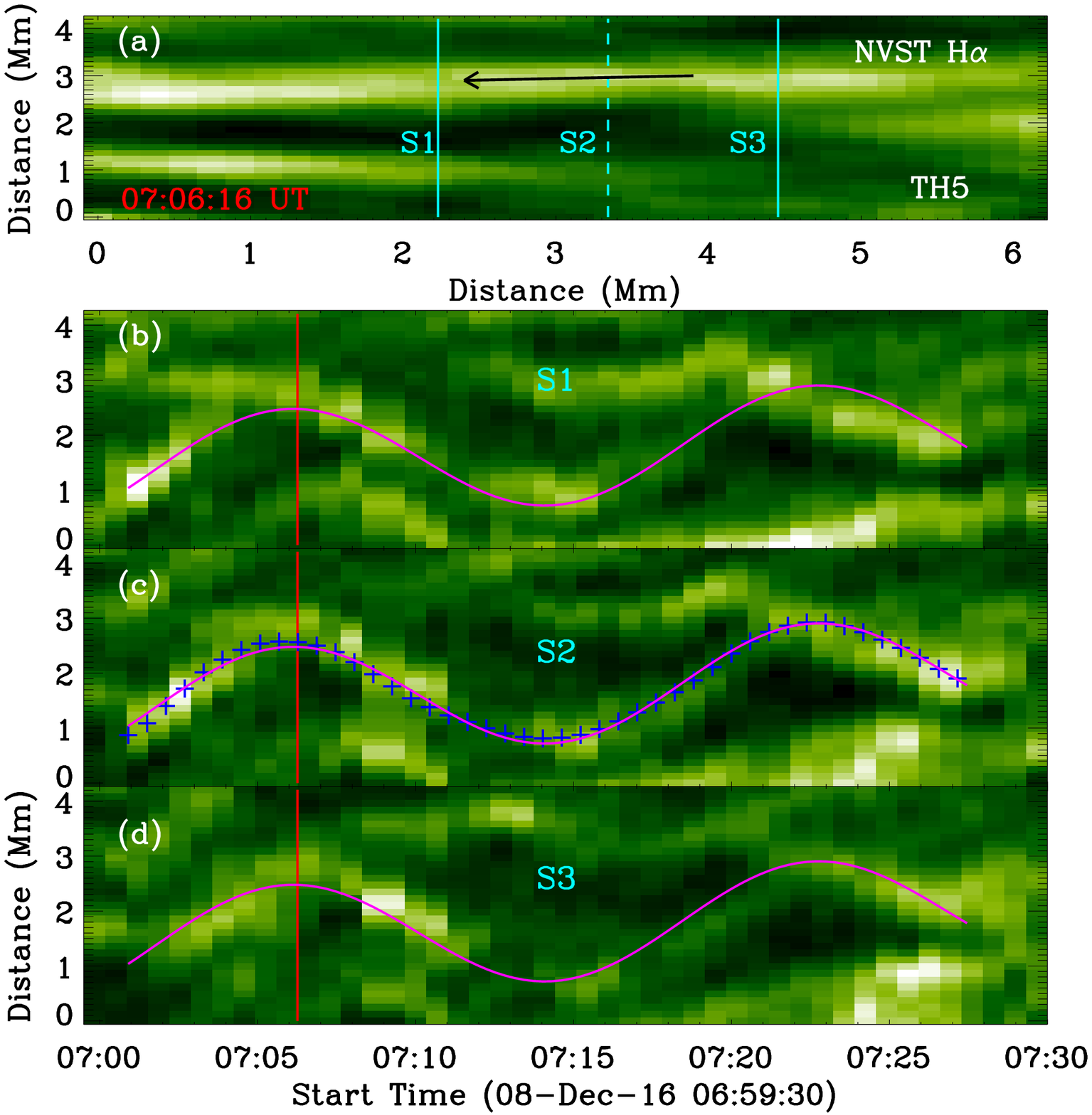}
\caption{Similar to \cref{slit1} but for the prominence thread TH5.
\label{slit5}}
\end{figure}

\begin{figure}[H]
\centering
\includegraphics[width=0.45\textwidth]{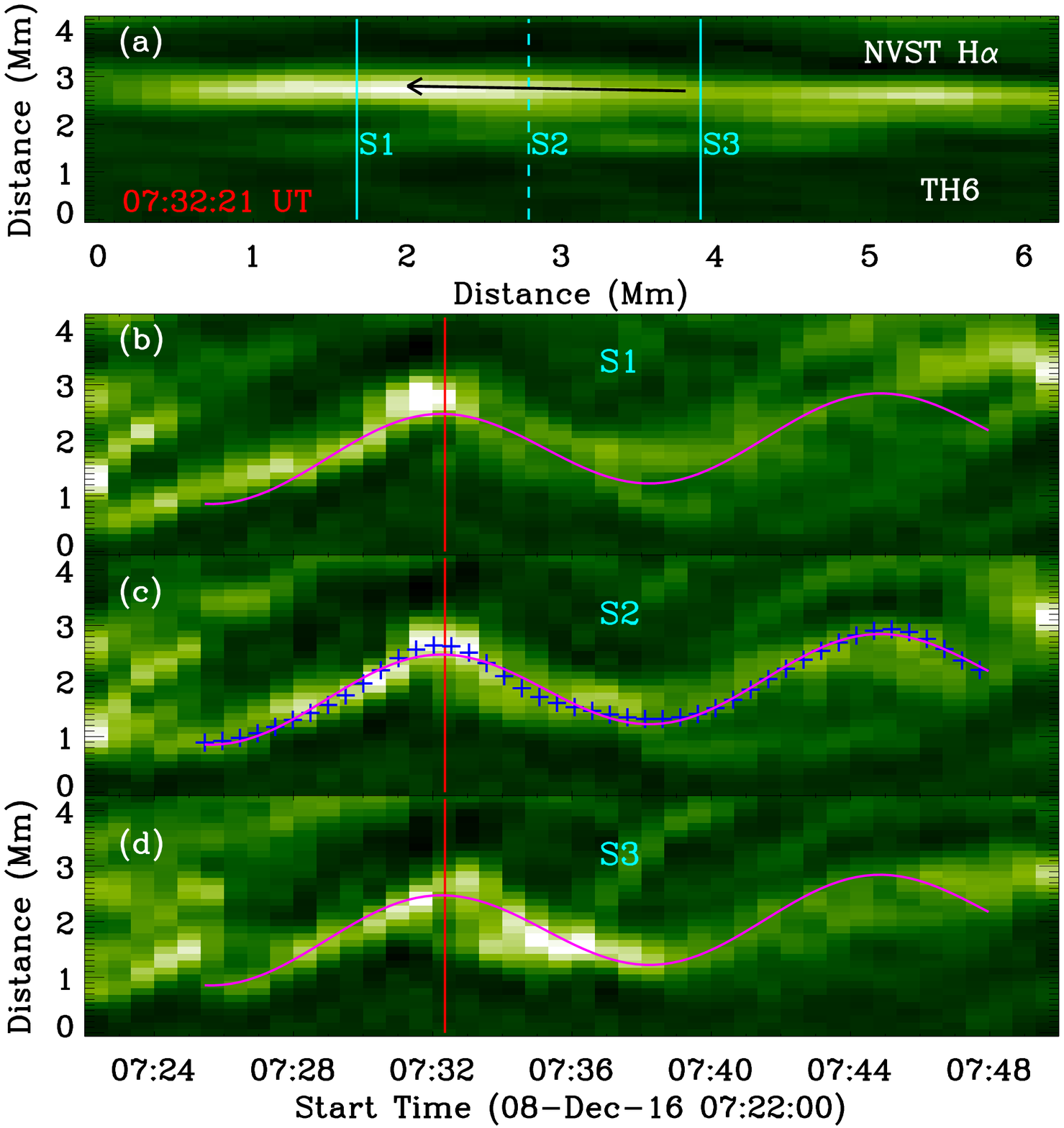}
\caption{Similar to \cref{slit1} but for the prominence thread TH6.
\label{slit6}}
\end{figure}

\section{Prominence seismology}
The upper limit of an uncertainty in the phase of the transverse
oscillation is estimated to be the ratio of the NVST time cadence
and the detected oscillatory period ($P$). This is because we cannot
find any significantly phase shift (or time delay) at different cut
slit locations along the entire length of each moving thread,
suggesting that the time delay is much shorter than the time
resolution of the NVST. Then we can deduce the minimum wavelength
($\lambda_{min}$) of the transverse oscillation for a specific
oscillatory period ($P$) in the moving thread. The thread lengths
are measured from the H$\alpha$ images, regarding the lower time
resolution (i.e., $\sim$40~s) of the NVST, and the fine-scale thread
is moving quickly (see the animation of S1), the thread length
measured here is also a lower limit \cite{Okamoto07}.

Next, we can estimate the lower limit of phase/wave speed ($c_{ph}$)
by using the relationship between the minimum wavelength and period,
i.e., Eq.~\ref{eq2}, if we considered the transverse oscillation as
the transverse MHD wave in the solar corona \cite{Okamoto07,Lin09}.
Then, using Eqs.~\ref{eq3} and \ref{eq4}, the Alfv\'{e}n speed
($V_A$) and the implied magnetic field strength ($B$) of the
transverse MHD wave can also be calculated, basing on the kink model
in the solar atmosphere \cite{Tian12,Yuan16a,Yuan16b}.

\begin{align}
  c_{ph}~\approx~\frac{\lambda_{min}}{P}. \label{eq2} \\
  V_A~=~\frac{c_{ph}}{\sqrt{2}}. \label{eq3} \\
  B~\approx~V_A~\sqrt{\mu_0\rho}. \label{eq4}
\end{align}
where $\mu_0$ is the magnetic permittivity of free space, which is
$\mu_0=4\pi~\times~10^{-7}$~N~A$^{-2}$. While $\rho$ represents the
plasma density in the quiescent prominence. In this study, a
relatively low value of the plasma density \cite{Okamoto07,Lin09} in
the quiescent prominence is used as $\rho~=~1.67~\times~10^{-11}$ kg
m$^{-3}$ ($\sim$10$^{10}$~cm$^{-3}$), since all the estimations are
minimum estimations.

Now, let us estimate the wave energy flux density of the detected
kink MHD waves. Using Eqs.\ref{eq5} and \ref{eq6}, the time-averaged
wave energy flux ($<E_k>$) carried by the fast kink MHD wave can be
estimated in the moving thread of the quiescent prominence, as Ref.
\cite{Morton12,Goossens13,Yuan16a}.
\begin{align}
  <E_k>~\approx~\frac{1}{4}c_{ph}~(\rho~v^{2}+\frac{b^2}{\mu_0}). \label{eq5} \\
  b~\approx~B~\frac{2\pi}{\lambda}A_m. \label{eq6}
\end{align}
Here, $v$ is the velocity amplitude of the transverse oscillation,
$b$ is the Lagrangian perturbation of the magnetic field. We want to
state that the Eq.~\ref{eq6} is an estimation with the Lagrangian
displacement vector, according to Eq.~(4) in Ref. \cite{Yuan16a}.

\end{appendix}

\end{multicols}
\end{document}